\documentclass[a4paper,conference]{IEEEtran}
\usepackage{graphicx}
 \usepackage{algorithm}
 \usepackage{algpseudocode}
\usepackage{multirow}

\usepackage{array}
\newcolumntype{C}[1]{>{\centering\arraybackslash}p{#1}}

\usepackage{tikz}

\usepackage{amssymb}

\usepackage{amsmath}

\newcommand{\xvect}{\mathbf{x}}
\newcommand{\yvect}{\mathbf{y}}
\newcommand{\zvect}{\mathbf{z}}

\newcommand{\fvect}{\mathbf{f}}

\newcommand{\zerovect}{\mathbf{0}}

\newcommand{\thetavect}{\boldsymbol{\theta}}

\newcommand{\muvect}{\boldsymbol{\mu}}

\newcommand{\norm}{\mathcal{N}}

\ifCLASSINFOpdf
\else
\fi

\usepackage[bookmarks=false]{hyperref}

\begin{document}
%









\title{Bayesian Inference for Gaussian Process Classifiers with Annealing and Pseudo-Marginal MCMC}

\author{\IEEEauthorblockN{Maurizio Filippone}
\IEEEauthorblockA{School of Computing Science, University of Glasgow\\
Email: maurizio.filippone@glasgow.ac.uk}
}


%


\maketitle

\begin{abstract}

Kernel methods have revolutionized the fields of pattern recognition and machine learning.
Their success, however, critically depends on the choice of kernel parameters.
Using Gaussian process (GP) classification as a working example, this paper focuses on Bayesian inference of covariance (kernel) parameters using Markov chain Monte Carlo (MCMC) methods.
The motivation is that, compared to standard optimization of kernel parameters, they have been systematically demonstrated to be superior in quantifying uncertainty in predictions.
Recently, the Pseudo-Marginal MCMC approach has been proposed as a practical inference tool for GP models.
In particular, it amounts in replacing the analytically intractable marginal likelihood by an unbiased estimate obtainable by approximate methods and importance sampling.
After discussing the potential drawbacks in employing importance sampling, this paper proposes the application of annealed importance sampling.
The results empirically demonstrate that compared to importance sampling, annealed importance sampling can reduce the variance of the estimate of the marginal likelihood exponentially in the number of data at a computational cost that scales only polynomially.
The results on real data demonstrate that employing annealed importance sampling in the Pseudo-Marginal MCMC approach represents a step forward in the development of fully automated exact inference engines for GP models.

\end{abstract}


%
\IEEEpeerreviewmaketitle

\section{Introduction}

Kernel methods have revolutionized the fields of pattern recognition and machine learning due to their nonlinear and nonparametric modeling capabilities~\cite{Shawe-Taylor04}.
Their success, however, critically depends on the choice of kernel parameters.
In applications where accurate quantification of uncertainty in predictions is of primary interest, it has been argued that optimization of kernel parameters may not be desirable, and that inference using Bayesian techniques represents a much more reliable alternative~\cite{FilipponeAOAS12,FilipponeIEEETPAMI14,Neal99,Rue09,Taylor12}.

This paper focuses in particular on the problem of inferring covariance (kernel) parameters of Gaussian Process classification models using Markov chain Monte Carlo (MCMC) techniques.
The choice of using GP classification as a working example is that they are formulated in probabilistic terms and are therefore particularly suitable candidates for carrying out Bayesian inference of their kernel parameters.
The choice of employing MCMC based inference techniques is that for general GP models and for general kernels they offer an all-purpose solution to do so up to a given precision~\cite{Flegal08}, as discussed in~\cite{Neal99,FilipponeML13,Neal93}.
The formulation of GP classifiers, and that of GP models in general, makes use of a set of latent variables $\fvect$ that are assumed to be distributed according to a GP prior with covariance parameterized by a set of parameters $\thetavect$.
The application of MCMC to directly draw samples form the posterior distribution over covariance parameters would require the evaluation of the so called marginal likelihood, namely the likelihood where latent variables are integrated out of the model, which is analytically intractable.

Recently, the Pseudo-Marginal (PM) MCMC approach has been proposed as a practical way to efficiently infer covariance parameters in Gaussian process classifiers exactly~\cite{FilipponeIEEETPAMI14}.
In this approach, computations do not rely on the actual marginal likelihood, but on an unbiased estimate obtained by approximate methods and Importance Sampling (IS).
While the sampling of covariance parameters using PM MCMC improves on previous approaches for inferring covariance parameters, a large variance in the estimate of the marginal likelihood can negatively impact the efficiency of the PM MCMC approach, making convergence slow and efficiency low.
In~\cite{FilipponeIEEETPAMI14}, IS was based on an importance distribution obtained by Gaussian approximations to the posterior over latent variables~\cite{Rasmussen06,Kuss05,Nickisch08}.
For certain values of the covariance parameters, the posterior over latent variables can be strongly non-Gaussian and the approximation can be poor, thus leading to a large variance in the IS estimate of the marginal likelihood~\cite{Kuss05}.
This effect is exacerbated by the dimensionality of the problem that makes the variance of IS grow exponentially large~\cite{Neal01}.
In the case of GP classification, estimating the marginal likelihood entails an integration in as many dimensions as the number of data, so this effect might be problematic in the case of large data sets.

This paper presents the application of Annealed Importance Sampling (AIS)~\cite{Neal01} to obtain a low-variance unbiased estimate of the marginal likelihood\footnote{The code to reproduce all the results in this paper can be found here:\\www.dcs.gla.ac.uk/$\sim$maurizio/pages/code\_ea\_mcmc\_ais/}.
This paper empirically demonstrate that compared to IS, AIS can reduce the variance of the estimate of the marginal likelihood exponentially in the number of data at a computational cost that scales only polynomially.
Finally, two versions of PM MCMC approaches, employing AIS and IS respectively, are compared on five real data sets.
The results on these data demonstrate that employing AIS in the PM MCMC approach represents a step forward in the development of fully automated exact Bayesian inference engines for GP classifiers. 

The remainder of this paper is organized as follows.
Sections~\ref{sec:gp} and~\ref{sec:ea:gp} review GP models and their fully Bayesian treatment using the PM MCMC approach.
Section~\ref{sec:ais} presents AIS to obtain an unbiased estimate of the marginal likelihood in GP models that can be used in the PM MCMC approach.
Section~\ref{sec:results} reports results on synthetic and real data, and section~\ref{sec:conclusions} reports the conclusions.

\section{Bayesian Inference for GP Classification} \label{sec:gp}

Let $X = \{\xvect_1, \ldots, \xvect_n\}$ be a set of $n$ input data where $\xvect_i \in R^{d}$, and let $\yvect = \{y_1, \ldots, y_n\}$ be a set of associated observed binary responses $y_i \in \{-1, +1\}$.
GP classification models are a class of hierarchical models where labels $\yvect$ are modeled as being independently distributed according to a Bernoulli distribution.
The probability of class $+1$ for an input $\xvect_i$ is based on a latent variable $f_i$ and is defined as $p(y_i=+1 | f_i) = \Phi(f_i)$, where $\Phi$ is the cumulative distribution function of the standard normal distribution, so that $p(\yvect | \fvect) = \prod_{i = 1}^n \Phi(y_i f_i)$. 
Latent variables $\fvect = \{f_1, \ldots, f_n\}$ are assumed to be distributed according to a GP prior, where a GP is a set of random variables characterized by the fact that any finite subset of them is jointly Gaussian.
GPs are specified by a mean function and a covariance function; for the sake of simplicity, in the remainder of this paper we will employ zero mean GPs.
The covariance function $k(\xvect, \xvect^{\prime} | \thetavect)$ gives the covariance between latent variables at inputs $\xvect$ and $\xvect^{\prime}$ and it is assumed to be parameterized by a set of parameters $\thetavect$.
This specification results in a multivariate Gaussian prior over the latent variables $p(\fvect | \thetavect) = \norm(\fvect | \zerovect, K)$ with $K$ defined as an $n \times n$ matrix with entries $k_{ij} = k(\xvect_i, \xvect_j | \thetavect)$.

A GP can be viewed as a prior over functions and it is appealing in situations where it is difficult to specify a parametric form for the function mapping $X$ into the probabilities of class labels. 
The covariance plays the role of the kernel in kernel machines, and in the remainder of this paper it will be assumed to be the Radial Basis Function (RBF) covariance
\begin{equation} \label{eq:rbf:covariance}
k(\xvect_i, \xvect_j | \thetavect) = 
\sigma \exp\left[-\frac{1}{2} \sum_{r=1}^d \frac{ (x_{ir} - x_{jr})^2}{\tau_r^2} \right].
\end{equation}
There can be one length-scale parameters $\tau_r$ for each feature, which is a suitable modelling assumption for Automatic Relevance Determination (ARD)~\cite{Mackay94}, or there can be one global length-scale parameter $\tau$ such that $\tau_1 = \ldots = \tau_d = \tau$.
The parameter $\sigma$ represents the variance of the marginal distribution of each latent variable.
A complete specification of a fully Bayesian GP classifier requires a prior $p(\thetavect)$ over $\thetavect$. 

When predicting the label $y_*$ for a new input data $\xvect_*$, it is necessary to estimate or infer all unobserved quantities in the model, namely $\fvect$ and $\thetavect$.
An appealing way of calculating predictive distributions is as follows:
\begin{equation} \label{eq:predictive}
p(y_* | \yvect) = \int p(y_* | f_*) p(f_* | \fvect, \thetavect) p(\fvect, \thetavect | \yvect) df_* d\fvect d\thetavect.
\end{equation}
In the last expression predictions are no longer conditioned on latent variables and covariance parameters, as they are integrated out of the model.
Crucially, such an integration accounts for the uncertainty in latent variables and covariance parameters based on their posterior distribution $p(\fvect, \thetavect | \yvect)$.

In order to compute the predictive distribution in eq.~\ref{eq:predictive}, a standard way to proceed is to approximate it using a Monte Carlo estimate:
\begin{equation} \label{eq:monte:carlo:integration}
p(y_* | \yvect) \simeq \frac{1}{N} \sum_{i=1}^N \int p(y_* | f_*) p(f_* | \fvect^{(i)}, \thetavect^{(i)}) df_* ,
\end{equation}
provided that samples from the posterior $p(\fvect, \thetavect | \yvect)$ are available.
Note that in the case of GP classification, the remaining integral has a closed form solution~\cite{Rasmussen06}.

As it is not possible to directly draw samples from $p(\fvect, \thetavect | \yvect)$, alternative ways to characterize it have been proposed.
A popular way to do so employs deterministic approximations to integrate out latent variables~\cite{Kuss05,Nickisch08}, but there is no way to quantify the error introduced by these approximation.
Also, quadrature is usually employed to integrate out covariance parameters, thus limiting the applicability of GP models to problems with few covariance parameters~\cite{Rue09}.
Such limitations might not be acceptable in some pattern recognition applications, so we propose Markov chain Monte Carlo (MCMC) based inference as a general framework for tackling inference problems exactly in GP models.
The idea underpinning MCMC methods for GP models is to set up a Markov chain with $p(\fvect, \thetavect | \yvect)$ as invariant distribution. 

To date, most MCMC approaches applied to GP models alternate updates of latent variables and covariance parameters.
All these approaches, however, face the complexity of having to decouple latent variables and covariance parameters, whose posterior dependence makes convergence to the posterior distribution slow.
Reparameterization techniques are a popular way to attempt to decouple the two groups of variables~\cite{Murray10,Papaspiliopoulos07,Yu11}.
Also, jointly sampling $\fvect$ and $\thetavect$ has been attempted in \cite{KnorrHeld02,Rue04}, and it is based on approximations to the posterior over latent variables.
Despite these efforts, a satisfactory way of sampling the parameters $\thetavect$ for general GP models is still missing, as demonstrated in a recent comparative study~\cite{FilipponeML13}. 

At this point it is useful to notice that samples from the posterior distribution of latent variables and covariance parameters can be obtained by alternating the sampling from $p(\fvect | \thetavect, \yvect)$ and $p(\thetavect | \yvect)$.
Obtaining samples from $p(\thetavect | \yvect)$ is obviously difficult, as it requires the marginal likelihood $p(\yvect | \thetavect)$; except for the case of a Gaussian likelihood, evaluating the marginal likelihood entails an integration which cannot be computed analytically~\cite{Rasmussen06}.
In the next section we will focus on the PM MCMC approach as a practical way of dealing with this problem.

Obtaining samples from $p(\fvect | \yvect, \thetavect)$, instead, can be done efficiently using Elliptical Slice Sampling (Ell-SS)~\cite{Murray10b}.
Ell-SS defines a transition operator $T(\fvect^{\prime} | \fvect)$, and is a variant of Slice Sampling~\cite{Neal03} adapted to the sampling of latent variables in GP models. 
Ell-SS begins by randomly choosing a threshold $\eta$ for $\log[p(\yvect | \fvect)]$
\begin{equation}
u \sim U[0,1] \qquad \eta = \log[p(\yvect | \fvect)] + \log[u]
\end{equation}
and by drawing a set of latent variables $\zvect$ from the prior $\norm(\mathbf{0}, K)$.
Then, a combination of $\fvect$ and $\zvect$ is sought, such that the log-likelihood of the resulting combination is larger than the threshold $\eta$.
Such a combination is defined by means of sine and cosine of an auxiliary variable $\alpha$, which makes the resulting combination spanning a domain of points that is an ellipse in the latent variable space.
The search procedure is based on slice sampling on $\alpha$ starting from the interval $[0, 2\pi]$.
Due to the fact that Ell-SS does not require any tuning and it has been shown to be very efficient for several GP models~\cite{FilipponeML13}, it is the operator that will be used in the remainder of this paper to sample latent variables.
However, note that latent variables can be also efficiently sampled by means of a variant of Hybrid Monte Carlo~\cite{FilipponeML13}.
\section{Pseudo-Marginal Inference for GP models} \label{sec:ea:gp}

For the sake of simplicity, this work will focus on the Metropolis-Hastings (MH) algorithm~\cite{Neal93,Hastings70} to obtain samples from the posterior distribution over covariance parameters.
The MH algorithm is based on the iteration of the following two steps: (i) proposing a new set of parameters $\thetavect^{\prime}$ drawing from a user defined proposal distribution $\pi(\thetavect^{\prime} | \thetavect)$ and (ii) evaluating the Hastings ratio
\begin{equation}
\tilde{z} = \frac{p(\yvect | \thetavect^{\prime}) p(\thetavect^{\prime})}{p(\yvect | \thetavect) p(\thetavect)} \frac{\pi(\thetavect | \thetavect^{\prime})}{\pi(\thetavect^{\prime} | \thetavect)}
\end{equation}
to accept or reject $\thetavect^{\prime}$.
As previously discussed, the marginal likelihood $p(\yvect | \thetavect) = \int p(\yvect | \fvect) p(\fvect | \thetavect) d\fvect$ cannot be computed analytically, except for the case of a Gaussian likelihood. 

The PM approach in~\cite{FilipponeIEEETPAMI14} builds upon a remarkable theoretical result~\cite{Beaumont03,Andrieu09} stating that it is possible to plug an unbiased estimate of the marginal likelihood $\tilde{p}(\yvect | \thetavect)$ in the Hastings ratio
\begin{equation}
\tilde{z} = \frac{\tilde{p}(\yvect | \thetavect^{\prime}) p(\thetavect^{\prime})}{\tilde{p}(\yvect | \thetavect) p(\thetavect)} \frac{\pi(\thetavect | \thetavect^{\prime})}{\pi(\thetavect^{\prime} | \thetavect)}
\end{equation}
and still obtain an MCMC algorithm sampling from the correct posterior distribution $p(\thetavect | \yvect)$.
In~\cite{FilipponeIEEETPAMI14} an unbiased estimate of the marginal likelihood was obtained as follows.
First, an approximation of the posterior over latent variables $p(\fvect | \yvect, \thetavect)$, say $q(\fvect | \thetavect, \yvect)$, was obtained by means of approximate methods, such as for example the Laplace Approximation (LA) or Expectation Propagation.
Second, based on $q(\fvect | \thetavect, \yvect)$, it was proposed to get an unbiased estimate of the marginal likelihood $p(\yvect | \thetavect)$ using IS.
In particular, this was achieved by drawing $N_{\mathrm{imp}}$ samples $\fvect^{(i)}$ from the approximating distribution $q(\fvect | \thetavect, \yvect)$.
Defining 
\begin{equation} \label{eq:importance:weights}
 w_{\mathrm{IS}}^{(i)} = \frac{p(\yvect | \fvect^{(i)}) p(\fvect^{(i)} | \thetavect)}{q(\fvect^{(i)} | \thetavect, \yvect)},
\end{equation}
the marginal likelihood $p(\yvect | \thetavect)$ was approximated by
\begin{equation} \label{eq:importance:pseudo}
\tilde{p}(\yvect | \thetavect) \simeq 
\frac{1}{N_{\mathrm{imp}}} \sum_{i=1}^{N_{\mathrm{imp}}} w_{\mathrm{IS}}^{(i)}.
\end{equation}
Such an estimate is unbiased and the closer $q(\fvect | \thetavect, \yvect)$ is to $p(\yvect | \fvect) p(\fvect | \thetavect)$ the lower the variance of the estimate~\cite{Neal01}.

In the experiments shown in \cite{FilipponeIEEETPAMI14} this estimate was adequate for the problems that were analyzed, especially when accurate approximations based on Expectation Propagation were used.
However, the variance of the IS estimate grows exponentially with the dimensionality of the integral~\cite{Neal01}, and this might represent a limitation when applying PM MCMC to large data sets.
In particular, a large variance in the estimate of $p(\yvect | \thetavect)$ can eventually lead to the acceptance of a $\thetavect$ because the corresponding marginal likelihood is overestimated.
If the overestimation is severe, it is unlikely that any new proposal will be accepted, resulting in slow convergence and low efficiency.
The aim of this paper is to present a methodology based on AIS~\cite{Neal01} which is capable of mitigating this effect.

\section{Marginal Likelihood estimation with Annealed Importance Sampling} \label{sec:ais}
\begin{figure*}[ht]
\begin{center}
{\scriptsize{\bf Annealing from the prior}} \\
\vspace{-2mm}\includegraphics[width=1\linewidth]{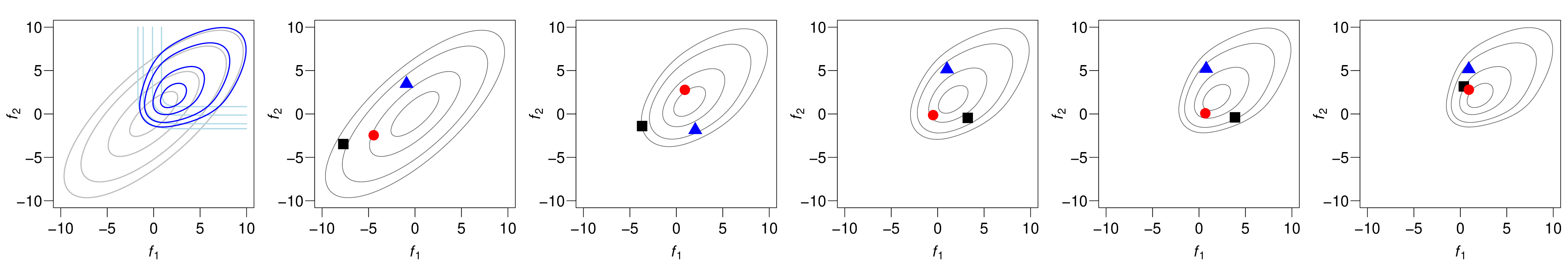} \\
{\scriptsize{\bf Annealing from an approximating distribution}} \\
\vspace{-2mm}\includegraphics[width=1\linewidth]{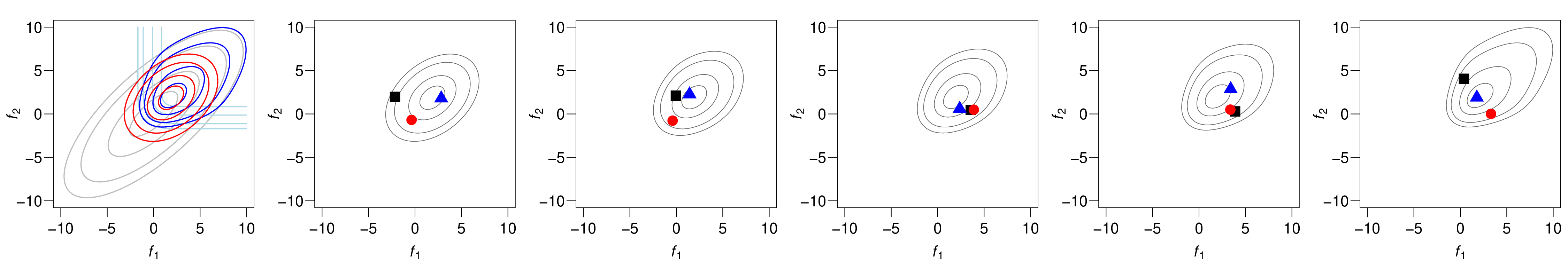}
\caption{Illustration of the annealing strategies studied in this work.
The figure was generated as follows. 
The input data $X$ comprises two data points in two dimensions with $\xvect_1=(-1,-1)$ and $\xvect_2 = (1,1)$, and corresponding labels $\yvect = (1,1)$.
The covariance is the one in eq.~\ref{eq:rbf:covariance} with $\sigma=15$ and $\tau=\exp(-1)$.
The leftmost plots show the multiplication of the GP prior (grey) and the likelihood (light blue) resulting in the posterior distribution over the two latent variables (blue).
The first row of the figure shows the annealing procedure from the GP prior to the posterior.
The leftmost plot in the second row shows prior, likelihood and posterior as before, along with the Gaussian approximation given by the LA algorithm (red).
The remaining plots in the second row show the annealing procedure from the approximating Gaussian distribution to the posterior.
In both cases, we defined $\beta_j = \exp(-j/2)$, thus assuming a geometric spacing for the $\beta$'s.
Three samples drawn from $g_s(\fvect)$ and propagated using operators $T_{i}(\fvect^{\prime} | \fvect)$ (one iteration of Ell-SS) have also been added to the plots. 
}
\label{fig:explain:annealing}
\end{center}
\end{figure*}

AIS is an extension of IS where the weights in eq.~\ref{eq:importance:weights} are computed based on a sequence of distributions going from one that is easy to sample from to the posterior distribution of interest. 
Following the derivation in~\cite{Neal01}, 
define $g_s(\fvect)$ as the unnormalized density of a distribution which is easy to sample from; in the next section we will study two of such distributions.
Also, define
\begin{equation}
g_0(\fvect) = p(\yvect | \fvect) p(\fvect | \thetavect) \propto p(\fvect | \thetavect, \yvect).
\end{equation}
AIS defines a sequence of intermediate unnormalized distributions
\begin{equation}
g_j(\fvect) = g_0(\fvect)^{\beta_j} g_s(\fvect)^{1 - \beta_j}
\end{equation}
with $1 = \beta_0 > \ldots > \beta_s = 0$.
The AIS sampling procedure begins by drawing one sample $\fvect_{s-1}$ from $g_s(\fvect)$.
After that, for $i = s-1, \ldots, 1$, a new $\fvect_{i-1}$ is obtained from $\fvect_{i}$ by iterating a transition operator $T_{i}(\fvect^{\prime} | \fvect)$ that leaves the normalized version of $g_i(\fvect)$ invariant.
Finally, computing the average of the following weights
\begin{equation}
w_{\mathrm{AIS}}^{(i)} = \frac{g_{s-1}(\fvect_{s-1})}{g_{s}(\fvect_{s-1})} \frac{g_{s-2}(\fvect_{s-2})}{g_{s-1}(\fvect_{s-2})} \cdots \frac{g_{1}(\fvect_{1})}{g_{2}(\fvect_{1})} \frac{g_{0}(\fvect_{0})}{g_{1}(\fvect_{0})}
\end{equation}
yields an unbiased estimate of the ratio of the normalizing constants of $g_0(\fvect)$ and $g_s(\fvect)$, which immediately yields an unbiased estimate of $p(\yvect | \thetavect)$.
For numerical reasons, it is safe to implement the calculations using logarithm transformations. 
Also, note that although the annealing strategy is inherently serial, the computations with respect to multiple importance samples can be parallelized.
We now analyze two ways of implementing AIS for GP models, which are visually illustrated in fig.~\ref{fig:explain:annealing}.

\subsection{Annealing from the prior}
When annealing from the prior, the intermediate distributions are between $g_s(\fvect) = \norm(\fvect | 0, K)$ and $g_0(\fvect) = \norm(\fvect | 0, K) p(\yvect | \fvect)$, namely
\begin{equation}
g_j(\fvect) = \norm(\fvect | 0, K) \left[p(\yvect | \fvect)\right]^{\beta_j}.
\end{equation}
Employing Ell-SS as a transition operator for $\fvect$ for the intermediate unnormalized distributions $g_j(\fvect)$ is straightforward, as the log-likelihood is simply scaled by $\beta_j$.
Annealing from the prior was proposed in~\cite{Kuss05} where it was reported that a sequence of $8000$ annealed distributions was employed.
This is because the prior and the posterior look very much different (see fig.~\ref{fig:explain:annealing}) and the only way to ensure a smooth transition from the prior to the posterior is by using several intermediate distributions.
This is problematic from a computational perspective, as the calculation of the marginal likelihood has to be done at each iteration of the PM approach to sample from the posterior distribution over $\thetavect$.
We therefore propose an alternative starting distribution $g_s(\fvect)$ that leads to a reduction in the number of intermediate distributions while obtaining estimates of the marginal likelihood that are accurate enough to ensure good sampling efficiency when used in the PM MCMC approach.

\subsection{Annealing from an approximating distribution}
Several Gaussian-based approximation schemes to integrate out latent variables have been proposed for GP models~\cite{Minka01,Opper00}.
When an approximation to the posterior over latent variables is available, it might be reasonable to construct the sequence of intermediate distributions in AIS starting from it rather than the prior.
When annealing from an approximating Gaussian distribution, the intermediate distributions are between $g_s(\fvect) = q(\fvect | \thetavect, \yvect) = \norm(\fvect | \muvect, \Sigma)$ and $g_0(\fvect) = \norm(\fvect | 0, K) p(\yvect | \fvect)$.
In order to employ Ell-SS as a transition operator $T_i(\fvect^{\prime} | \fvect)$, it is useful to write the unnormalized intermediate distributions as
\begin{equation}
g_j(\fvect) = \norm(\fvect | \muvect, \Sigma) \left[\frac{\norm(\fvect | 0, K) p(\yvect | \fvect)}{\norm(\fvect | \muvect, \Sigma)}\right]^{\beta_j}.
\end{equation}
In this way, the model can be interpreted as having a prior $\norm(\fvect | \muvect, \Sigma)$ and a likelihood given by the term in square brackets; 
applying Ell-SS to this formulation is straightforward.

\section{Experimental results} \label{sec:results}

The first part of this section, compares the behavior of IS and AIS in the case of synthetic data.
The second part of this section, reports an analysis of IS and AIS when employed in the PM MCMC approach applied to real data.
In all experiments, the approximation was based on the Laplace Approximation (LA) algorithm.
Also, we imposed Gamma priors on the parameters $\mathrm{Ga}(\sigma | a=1.1, b=0.1)$ and $\mathrm{Ga}(\tau_i | a=1, b=1)$ for the ARD covariance and $\mathrm{Ga}(\tau | a=1, b=1/\sqrt{d})$ for the isotropic covariance, where $a$ and $b$ are shape and rate parameters respectively.
Following the recommendations in~\cite{Neal01,Neal96b}, $s = \sqrt{n}$ intermediate distributions were defined based on a geometric spacing of the $\beta$'s.
In particular, this was implemented by setting $s/2-1$ uniformly spaced values of $\log[\beta]$ between $\log[1]$ and $\log[0.2]$, $s/2$ uniformly spaced values between $\log[0.2]$ and $\log\left[10^{-6}\right]$, and finally $\beta_s = 0$.
In AIS, the transitions $T_{i}(\fvect^{\prime} | \fvect)$ involved one iteration of Ell-SS.

\subsection{Synthetic data}

The aim of this section is to highlight the potential inefficiency in employing IS to obtain an unbiased estimate of the marginal likelihood and to demonstrate the effectiveness of AIS in dealing with this problem.
In particular, this can be problematic in large dimensions, namely when analyzing large amounts of data.
In order to show this effect, we generated data sets with an increasing number of data $n=10, 50, 100, 500, 1000$ in two dimensions with a balanced class distribution.
Data were generated drawing input vectors uniformly in the unit square and a latent function from a GP with covariance in eq.~\ref{eq:rbf:covariance} with $\sigma=20$ and a global $\tau=0.255$.
This combination of covariance parameters leads to a strongly non-Gaussian posterior distribution over the latent variables making IS perform poorly when $n$ is large.

\begin{figure}[t]
\begin{center}
 \includegraphics[width=0.98\linewidth]{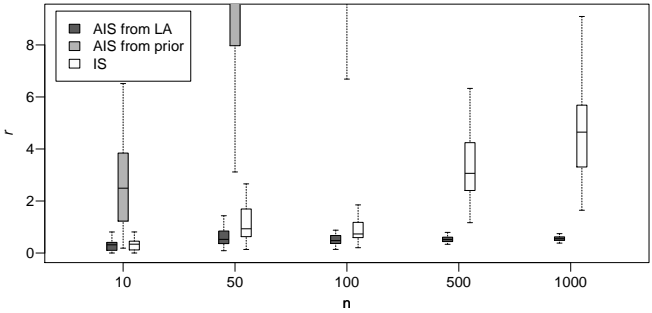} \\
\caption{This figure shows a measure of the quality of the IS and AIS (annealing from the prior and from an approximating distribution obtained by the LA algorithm) estimators of the marginal likelihood.
The boxplot summarizes the distribution of $r$ in eq.~\ref{eq:r:score} for $50$ values of $\thetavect$ drawn from $p(\thetavect | \yvect)$.
}
\label{fig:res:synth}
\end{center}
\end{figure}

In order to obtain a measure of variability of the IS and AIS estimators of the marginal likelihood, we analyze the standard deviation of the estimator of $\log[p(\yvect | \thetavect)]$
\begin{equation} \label{eq:r:score}
r = \mathrm{st\ dev} \left\{ \log_{10} \left[ \tilde{p}(\yvect | \thetavect) \right] \right\}.
\end{equation}
In the experiments, $r$ was estimated based on $50$ repetitions; fig.~\ref{fig:res:synth} shows the distribution of $r$ based on $50$ draws of $\thetavect$ from the posterior $p(\thetavect | \yvect)$ obtained from a preliminary run of an MCMC algorithm.
Ideally, a perfect estimator of the marginal likelihood would yield a degenerate distribution of $r$ over posterior samples of $\thetavect$ at zero.
In practice, the distribution of $r$ indicates the variability (across posterior samples of $\thetavect$) around an average value of the standard deviation of the estimator of the logarithm of the marginal likelihood.
The representation in $\log_{10}$ is helpful to get an idea of the order of magnitude of such a variability.
For instance, a distribution of $r$ across posterior samples of $\thetavect$ concentrated around $2$ would mean that, on average, the estimates of the marginal likelihood span roughly two orders of magnitude.

Fig.~\ref{fig:res:synth} shows the distribution of $r$ for AIS when annealing from the prior and from an approximating distribution, along with the distribution of $r$ for IS as in~\cite{FilipponeIEEETPAMI14}.
In all methods we set $N_{\mathrm{imp}} = 4$.
The results confirm that annealing from the prior offers much poorer estimates of the marginal likelihood compared to annealing from an approximating distribution and will not be considered further.
The analysis of the results in fig.~\ref{fig:res:synth} reveal that when annealing from an approximating distribution, the reduction in variance of the estimate of the marginal likelihood compared to IS is exponential in $n$.
When comparing the computational cost of running IS and AIS, instead, we notice that AIS increases it by a factor which scales only polynomially with $n$.
This is because, after approximating the posterior over $\fvect$ (that typically costs $O(n^3)$ operations), 
in AIS drawing the initial importance samples, iterating Ell-SS, and computing the weights $w_{\mathrm{AIS}}$ costs $O(n^2)$ operations; this needs to be done as many times as the number of intermediate distributions $s$, which in our case means $O(\sqrt{n})$ times.
In IS, drawing the importance samples and computing the weights $w_{\mathrm{IS}}$ requires $O(n^2)$ operations.

\subsection{Real data}

\begin{table*}[th]
\renewcommand{\arraystretch}{1.3}
\caption{Comparison between the average acceptance rate (in \%) obtained by the PM MCMC approach using IS and AIS. 
The number in parentheses represents the standard deviation of the average acceptance rate across five parallel chains.
}
\label{tab:res:real}
\centering
\begin{tabular}{|c|C{1.25cm}|C{1.25cm}|C{1.25cm}|C{1.25cm}|C{1.25cm}|C{1.25cm}|C{1.25cm}|C{1.25cm}|C{1.25cm}|C{1.25cm}|}
\multicolumn{11}{c}{ {\bf Isotropic covariance} } \\ 
\hline 
 & \multicolumn{2}{c|}{ Glass } & \multicolumn{2}{c|}{ Thyroid } & \multicolumn{2}{c|}{ Breast } & \multicolumn{2}{c|}{ Pima } & \multicolumn{2}{c|}{ Banknote }\\ 
 & \multicolumn{2}{c|}{$n = 214$, $d = 9$} & \multicolumn{2}{c|}{$n = 215$, $d = 5$} & \multicolumn{2}{c|}{$n = 682$, $d = 9$} & \multicolumn{2}{c|}{$n = 768$, $d = 8$} & \multicolumn{2}{c|}{$n = 1372$, $d = 4$}\\ 
\cline{2-11} 
$N_{\mathrm{imp}}$ & IS & AIS & IS & AIS & IS & AIS & IS & AIS & IS & AIS\\ 
\hline 
1 & $2.8 (1.6)$ & $5.2 (1.9)$ & $1.1 (1.0)$ & $3.2 (2.3)$ & $17.9 (2.4)$ & $28.0 (2.7)$ & $24.8 (1.4)$ & $29.3 (2.6)$ & $1.1 (0.6)$ & $3.2 (3.9)$\\ 
10 & $10.4 (3.1)$ & $11.4 (5.3)$ & $4.1 (3.8)$ & $6.4 (3.9)$ & $30.5 (4.1)$ & $36.4 (3.5)$ & $30.8 (2.6)$ & $30.8 (1.7)$ & $4.7 (1.0)$ & $9.2 (5.6)$\\ 
\hline 
\end{tabular}
\\ \vspace{1mm}
\begin{tabular}{|c|C{1.25cm}|C{1.25cm}|C{1.25cm}|C{1.25cm}|C{1.25cm}|C{1.25cm}|C{1.25cm}|C{1.25cm}|C{1.25cm}|C{1.25cm}|}
\multicolumn{11}{c}{ {\bf ARD covariance} } \\ 
\hline 
 & \multicolumn{2}{c|}{ Glass } & \multicolumn{2}{c|}{ Thyroid } & \multicolumn{2}{c|}{ Breast } & \multicolumn{2}{c|}{ Pima } & \multicolumn{2}{c|}{ Banknote }\\ 
 & \multicolumn{2}{c|}{$n = 214$, $d = 9$} & \multicolumn{2}{c|}{$n = 215$, $d = 5$} & \multicolumn{2}{c|}{$n = 682$, $d = 9$} & \multicolumn{2}{c|}{$n = 768$, $d = 8$} & \multicolumn{2}{c|}{$n = 1372$, $d = 4$}\\ 
\cline{2-11} 
$N_{\mathrm{imp}}$ & IS & AIS & IS & AIS & IS & AIS & IS & AIS & IS & AIS\\ 
\hline 
1 & $1.3 (1.3)$ & $3.6 (2.3)$ & $0.4 (0.3)$ & $2.9 (1.8)$ & $1.8 (1.7)$ & $5.0 (2.5)$ & $17.1 (2.7)$ & $22.5 (3.3)$ & $1.3 (1.4)$ & $4.7 (2.1)$\\ 
10 & $2.5 (1.6)$ & $4.9 (3.2)$ & $6.9 (2.4)$ & $6.4 (2.0)$ & $7.7 (2.6)$ & $4.5 (1.8)$ & $22.8 (4.0)$ & $24.1 (3.9)$ & $5.8 (3.3)$ & $9.2 (3.1)$\\ 
\hline 
\end{tabular}

\end{table*}

This section reports an analysis of the PM MCMC approach applied to five UCI data sets~\cite{Asuncion07} when the marginal likelihood is estimated using AIS and IS.
The Glass data set is multi-class, and we turned it into a two class data set by considering the data labelled as ``window glass'' as one class and data labelled as ``non-window glass'' as the other class.
In all data sets, features were normalized to have zero mean and unit standard deviation.
All experiments were repeated varying the number of importance samples $N_{\mathrm{imp}} = 1, 10$, and employing isotropic and ARD RBF covariance functions as in eq.~\ref{eq:rbf:covariance}.

In order to tune the MH proposal, we ran a preliminary MCMC algorithm for $2000$ iterations.
This was initialized from the prior and the marginal likelihood in the Hastings ratio was obtained by the LA algorithm.
The proposal was then adapted to obtain an acceptance rate between $20\%$ and $30\%$.
This set up was useful in order to avoid problems in tuning the proposal mechanism when a noisy version of the marginal likelihood is used, which may lead to a poor acceptance rate independently of the proposal mechanism.
Tab.~\ref{tab:res:real} reports the average acceptance rate when switching to an unbiased version of the marginal likelihood obtained by IS or AIS for different values of $N_{\mathrm{imp}}$ after the adaptive phase.
The average acceptance rate was computed based on $1500$ iterations, collected after discarding $500$ iterations, and over $5$ parallel chains.

The results are variable across data sets and the type of covariance, but the general trend is that employing AIS in the PM MCMC approach improves on the acceptance rate compared to IS.
In a few cases, it is striking to see how replacing an approximate marginal likelihood with an unbiased estimate in the Hastings ratio does not affect the acceptance rate, thus confirming the merits of the PM MCMC approach.
In general, however, PM MCMC is affected by the use of an estimate of the marginal likelihood.
In cases where this happens, AIS consistently offers a way to reduce the variance of the estimate of the marginal likelihood compared to IS, and this improves on the acceptance rate.



\section{Conclusions} \label{sec:conclusions}

This paper presented the application of annealed importance sampling to obtain an unbiased estimate of the marginal likelihood in GP classifiers.
Annealed importance sampling for GP classifiers was previously proposed in~\cite{Kuss05} where the sequence of distributions was constructed from the prior to the posterior over latent variables.
Given the difference between these two distributions, the annealing strategy requires the use of several intermediate distributions, thus making this methodology impractical.
This paper studied the possibility to construct a sequence of distributions from an approximating distribution rather than the prior, and empirically demonstrated that, compared to importance sampling, this reduces the variance of the estimator of the marginal likelihood exponentially in the number of data. 
Crucially, this reduction comes at a cost that is only polynomial in the number of data.
Also, annealed importance sampling can be easily parallelized.


The motivation for studying this problem was to plug the unbiased estimate of the marginal likelihood in the Hastings ratio in order to obtain an MCMC approach sampling from the correct posterior distribution over covariance parameters.
The results on real data show that employing importance sampling within the pseudo-marginal MCMC approach can be satisfactory in many cases.
However, in general, annealed importance sampling leads to a lower variance estimator of the marginal likelihood, and the resulting pseudo-marginal MCMC approach significantly improves on the average acceptance rate.
These results suggest a promising direction of research towards the development of MCMC methods where the likelihood is estimated in an unbiased fashion, but the acceptance rate is as if the likelihood were known exactly.
Given that the computational overhead scales with less than the third power of the number of data, the results indicate that this can be achieved with an acceptable computational cost.

This paper considered GP classification as a working example, and the Laplace approximation algorithm to obtain the importance distribution. 
A matter of current investigation is the application of the proposed methodology to other GP models and other approximation schemes.
Furthermore, this paper focused on the case of full covariance matrices.
These results can be extended to deal with sparse inverse covariance matrices, which are popular when modeling spatio-temporal data, thus leading to the possibility to process massive amounts of data due to the use of sparse algebra routines.
Finally, this paper did not attempt to optimize the annealing scheme, but it would be sensible to do so in order to minimize the variance of the annealed importance sampling estimator of the marginal likelihood~\cite{Behrens10}.

\renewcommand{\url}[1]{}



%


\end{document}